\documentstyle[prd,aps,epsf,floats,amsfonts,amssymb,amsmath]{revtex}

\begin{document}
 \renewcommand{\theequation}{\thesection.\arabic{equation}}
 \draft

\title{Matter-Antimatter Asymmetry Generated by Loop Quantum Gravity}

\author{Gaetano Lambiase$^{* \dagger ~ a}$ and Parampreet Singh$^{\ddagger ~ b}$}
\address{$^*$Dipartimento di Fisica "E.R. Caianiello"
 Universit\'a di Salerno, 84081 Baronissi (Sa), Italy.}
  \address{$^\dagger$INFN - Gruppo Collegato di Salerno, Italy.}
\address{$^\ddagger$Inter-University Center for Astronomy and Astrophysics,
Post Bag 4, Ganeshkhind, Pune 411007, India.}

\maketitle
\begin{abstract}
We show that Loop Quantum Gravity provides new mechanisms through
which observed matter-antimatter asymmetry in the Universe can
naturally arise at temperatures less than GUT scale. This is
enabled through the introduction of a new length scale ${\cal L}$,
much greater than Planck length ($l_P$), to obtain semi-classical weave
states in the theory.  This scale which depends on the momentum of
the particle modifies the dispersion relation for different
helicities of fermions and leads to lepton asymmetry.
\end{abstract}
\pacs{PACS No.: 04.60.Pp, 11.30.Fs, 98.80.Cq}

\section{Introduction}
\setcounter{equation}{0}

Many theories of Quantum Gravity are expected to bring non-trivial
modifications to the underlying spacetime near Planck scale.
 Loop Quantum Gravity, which is one of the candidate theories of
Quantum Gravity, predicts a discrete spectrum for geometrical
operators \cite{rovelli}. However, inaccessibility of Planck scale
in laboratories poses a challenge to test such predictions. It is
hence desired that a contact  be made with the classical world
through some semi-classical techniques. This might also open a
window to see the signatures of Quantum Gravity at the level of
effective theories which may differ from  conventional low energy
theories.

Loop Quantum Gravity which is based on the quantization of
spacetime itself, results in a polymer like structure of quantum
spacetime. The classical spacetime is a coarse grained form of
underlying discrete quantum spacetime and one of the important
issues in Loop Quantum Gravity is to understand the transition
from discrete quantum spacetime to smooth classical spacetime.
Though the low energy sector of Loop Quantum Gravity and the
transition to the classical spacetime is yet to be completely
understood, there have been some attempts in this direction to
obtain the semi-classical states in the theory which include
construction of weave states which can approximate 3-metrics
\cite{weave} and via coherent states which peak around classical
trajectories \cite{coherent}. For more discussion on related
issues we refer the reader to a recent review \cite{smolin}.

The coherent state approach to understand low energy sector is
based on finding quantum states which give minimum dispersion for
the observables in the theory, whereas the weave state approach
involves a new length scale ${\cal L} \gg l_p$ such that for
distances $d \ll {\cal L}$ polymer structure of quantum spacetime
becomes manifest and for $d \geq {\cal L}$ one recovers continuous
flat classical geometry. This approach was extended to study the
construction of weave states describing gravity coupled to massive
spin-1/2 Majorana fields in a series of important papers by
Alfaro, Morales-T\'ecotl, and Urrutia (AMU)
\cite{alfaroPRL,alfaroPRD,hugo}. This new length scale modifies
the dispersion relation for different helicities of fermions  and
breaks Lorentz invariance in the theory.

Apart from Loop Quantum Gravity, deformations of the Lorentz
invariance manifest by means of a slight deviation from the
standard dispersion relations of particles propagating in the
vacuum have been suggested in various ways, see for eg.
\cite{amelino,dharam}. The modifications to dispersion relation
may  arise if the underlying spacetime is non-commutative
\cite{nonc}. These theories which are characterized by a
non-commutativity parameter of dimensions of square of length, may
serve as a description for foamy structure of quantum spacetime.
Similar modifications have also been studied in the framework of
String theory \cite{kostelecky,ellis}. These approaches foresee a
dispersion relation in vacuo of particles of the form (we shall
use natural units $c=1=\hbar$)
 \begin{equation}\label{general}
 E^2\approx p^2+m^2+f\left(M,p\;l_{P}\right)\,,
 \end{equation}
where $f(x)$ is a model dependent function, $M$ fixes a
characteristic scale not necessarily determined by Planck length
$l_{P}\sim 10^{-19}$GeV$^{-1}$, and $p\,l_{P}\ll 1$. As a
consequence of Eq. (\ref{general}), the {\it quantum gravitational
medium} responds differently to the propagation of particles of
different energies.

With the breakdown of the Lorentz invariance in the theory, $CPT$
violation is expected and so new mechanisms to generate observed
matter-antimatter asymmetry in the Universe. The matter-antimatter
asymmetry in the Universe is conventionally understood, for
example, through baryogenesis processes occurring at GUT or
electroweak scales. However, most of these conventional mechanisms
to generate this asymmetry in Standard Model or extensions of it
run into one or another problem \cite{dine}, like inflation would
significantly dilute the asymmetry produced during GUT era. Any
low temperature mechanism to generate this asymmetry is hence
highly desirable. The origin of matter-antimatter asymmetry in the
Universe thus remains one of the unsolved puzzles whose resolution
may be possible through some new aspects of physics arising
through a fundamental theory. There have been earlier proposals
based on quantum gravity framework to generate matter-antimatter
asymmetry, like from primordial spacetime foam \cite{primord},
quantum gravity deformed uncertainty relations \cite{qgdur} and
string based scenarios \cite{bertolami}. In this letter, we would
like to present another interesting scenario arising out from
quantum gravity through weave states. We would show that weave
states of spin-1/2 fields provide a natural mechanism to generate
matter-antimatter asymmetry at temperatures of the order of
reheating temperature of inflation, far below the GUT temperature.

\section{Lepton Asymmetry in AMU formalism}
\setcounter{equation}{0}

In the weave state approach, the goal is to find a loop state
which approximates a classical geometry at a scale much larger
than $l_P$. A semi-classical weave state corresponding to Majorana
fermions is characterized by a scale length ${\cal L}$ such that
$l_P \ll {\cal L} \leq \lambda_D = 1/p$, where $\lambda_D$ is the
de Broglie wavelength of the fermion and $p$ its corresponding
momentum. For the Dirac equation with quantum corrections to be
properly defined on a continuous flat spacetime arising through
weave state construction, it is required that the scale length
${\cal L} \leq 1/p$ \cite{alfaroPRL}. Such a scale is known as
mobile scale \cite{hugo} which is different for different
fermionic species and the upper bound on ${\cal L}$ corresponding
to the weave state of a particular fermion is  set by the momentum
of that fermion. One may also treat this scale as a universal
scale and obtain bounds on it by observations
\cite{AlfaroPalma,AlfaroPalma2,gaetanoMPLA}, however, in this
letter we would restrict to the case of ${\cal L}$ as a mobile
scale.

The introduction of scale ${\cal L}$  leads to modifications in dispersion
relation which have been studied in various interesting contexts
\cite{stecker,urrutiaCBR,major}. Similar
phenomena have also been studied for photons
\cite{alfaroPRD,gleiser,gambini}. The dispersion relation with
leading order terms in $l_P$ and ${\cal L}$ can be written as
\cite{AlfaroPalma2}
\begin{equation}
E_{\pm}^2 = (1 + 2 \alpha) \, p^2 + \eta \, p^4 \pm 2 \, \lambda \, p + m^2
\end{equation}
with
\begin{equation}
\alpha = \kappa_1 \left(\frac{l_P}{{\cal L}}\right)^2, ~~ \eta = \kappa_3 \,
l_P^2, ~~ \lambda = \kappa_5 \left(\frac{l_P}{2 {\cal
L}^2}\right)~~
\end{equation}
where $\kappa_1, \kappa_3$ and $\kappa_5$ are of the order of
unity. Here $+$ and $-$ refer to two helicity states of the
fermion. We would specialize to the limiting case of Majorana
fermions with vanishingly small mass, $m \longrightarrow 0$. In
this way we can treat the fermions as Weyl particles. We would
further neglect $l_P^2$ terms in comparison to other dominating
terms in the above dispersion relation. Thus, the dispersion
relation can be rewritten as
\begin{equation} \label{disp-ferm}
E_{\pm}^2 =  p^2   \pm 2 \, \lambda \, p ~~.
\end{equation}
It should be noted that though the above modification to the dispersion
relation is of linear in momentum, the correction term effectively
behaves as the one cubic in momentum. This is because ${\cal L}$ is
a mobile scale and its upper bound scales as $1/p$. Thus, the above
correction, which is similar to other cubic in momentum modifications to
dispersion relation \cite{smolin}, dies out rapidly at low momenta..
We would now discuss the implications of this dispersion relation for
the case of neutrinos, where the helicity dispersion can be casted in terms
of the difference between energy levels of particle and antiparticle
states which leads to a net difference in their number densities
and hence lepton asymmetry in this framework.

Matter-antimatter asymmetry is generally understood through
Sakharov conditions who in his seminal paper \cite{sakharov},
showed that to generate the non-zero baryonic number to entropy
$\eta_B \sim (2.6 - 6.2)\times 10^{-10}$ from a baryonic symmetric
universe, the following requirements are necessary: 1) baryon
number processes violating in particle interactions; 2) $C$ and
$CP$ violation in order that processes generating $B$ are more
rapid with respect to $\bar{B}$; 3) out of the equilibrium: since
$m_B=m_{\bar{B}}$, as follows from $CPT$ symmetry, the equilibrium
space phase density of particles and antiparticles are the same.
To maintain the number of baryon and anti-baryon different, i.e.
$n_B\neq n_{\bar{B}}$, the reaction should freeze out before
particles and antiparticles achieve the thermodynamical
equilibrium.

GUT theories offer an ideal setting for Sakharov's conditions to
be satisfied \cite{kolb}. Baryon number violation occurs in these
theories since gauge bosons mediate interactions that transform
quarks into leptons and anti-quarks. $C$ is maximally violated in
the electro-weak sector, and $CP$ violation follows by making the
coupling constants of lepto-quark gauge bosons complex. Finally,
out of equilibrium condition is achieved by the expansion of the
universe when the reaction rates become lower than the Hubble
expansion rates at some {\it freeze-out} temperature. Such a
temperature is characterized by the decoupling temperature $T_d$,
which, in the GUT baryogenesis scenario, is given by $T_d\sim
10^{16}$GeV. However, GUT baryogenesis runs into problems because
inflation occurring at similar temperature dilutes the baryon
asymmetry. For baryons to be produced after inflation it is
necessary to reheat the Universe to the scale of $M_{GUT}$ which
is unrealistic in inflationary scenarios. In fact, bounds on
gravitino production give the reheating temperature $T_R$ of the
order of $10^8-10^{10}$ GeV \cite{tr}, whereas in SUSY inflation
models this may be raised to $10^{12}$ GeV \cite{susytr}.
Similarly, processes like electroweak baryogenesis and
leptogenesis suffer from problems like very small region of
parameter space which can yield asymmetry and lack of direct
measurement of relevant parameters \cite{dine}.

It is worth to quote some alternative mechanisms proposed in
literature which are not based on quantum gravity.
As observed in Refs. \cite{kaplan,dolgov}, if the $CPT$
symmetry and the baryon number is violated, a baryon asymmetry
could arise in thermal equilibrium. This mechanism to generate
baryon asymmetry has been applied in different contexts: the
spontaneous breaking of $CPT$ induced by the coupling of baryon
number current with a scalar field \cite{kaplan};  baryogenesis asymmetry
generated from primordial tensor perturbation \cite{subhendra} and
matter-antimatter asymmetry through interaction between gravitational
curvature and fermionic spin \cite{ms}.
For other mechanisms related to the lepton asymmetry, see Ref.
\cite{dolgovRep} and reference therein, as well as Refs.
\cite{allLeptAsym}.

In Loop Quantum Gravity, the different dispersion relations of
particles having different helicity determines a deviation from
thermal equilibrium between neutrinos and anti-neutrinos,
$n(\nu)\neq n({\bar \nu})$, where $n(\nu)$ and $n({\bar \nu})$ are
the number density of positive helicity neutrinos and negative
helicity antineutrinos, respectively. We would further assume that
there are no additional mechanisms  which give rise to neutrino
asymmetry. In such a case, the deviation from the chemical
equilibrium, which generates the baryon asymmetry, occurs only due
to Loop Quantum Gravity effects and the expansion of the universe.
If neutrinos are produced with energy $E$, then the dispersion
relation (\ref{disp-ferm}) gives
\begin{equation}\label{Lenergy}
E^{2} = p^2 + 2 \lambda \, p  ~~ \Longrightarrow ~~  p = \sqrt{E^2+\lambda^2} - \lambda\,,
\end{equation}
for neutrinos, and
\begin{equation}\label{Renergy}
E^{2} = p^2 - 2 \lambda \, p  ~~ \Longrightarrow  p = \sqrt{E^2+\lambda^2} + \lambda\,,
\end{equation}
for anti-neutrinos. Note that energy dispersion relation {\it
forbids} anti-neutrinos to have $p \, \in \, [0, 2 \lambda]$,
whereas no such restriction arises for neutrinos. This is purely a
Loop Quantum Gravity effect induced through quantum structure of
spacetime which seemingly favors one helicity over another.

The number density of neutrinos at the equilibrium for a given
temperature $T$ is (for $k_B=1$)
\begin{equation}\label{n(T)}
  n(\nu)=\frac{gT^3}{2\pi^2}\int_{0}^{\chi} dx
  \,\frac{x(\sqrt{x^2+z}-\sqrt{z})^2}{\sqrt{x^2+z}}\,\frac{1}{e^x+1}\,,
\end{equation}
where $\chi = 1/({\cal L} T)$, $x = E/T$, $z=(\lambda/T)^2$,
$T$ satisfies the relation ${\dot T}=-HT$, and $H={\dot a}/a$,
being $a(t)$ the scale factor of the
universe, \cite{dolgov}. The dot stands for the derivative with
respect to the cosmic time. The departure from the chemical
equilibrium caused by the expansion of the universe is encoded in
the quantity ${\cal F}=3Hn+{\dot n}$ \cite{dolgov}, which turns
out to be
 \[
 {\cal F}({\nu})=2\left(\frac{\lambda}{T}\right)^2HT^3
 \frac{d}{dz}\left[\frac{g}{2\pi^2}\int_0^{\chi}dx
  \,\frac{x(\sqrt{x^2+z}-\sqrt{z})^2}{\sqrt{x^2+z}}\,\frac{1}{e^x+1}\right]\,.
 \]
It vanishes as $\lambda=0$.\footnote{In absence of Loop Quantum
Gravity corrections, the deviation from the chemical equilibrium
occurs only if particles are massive. In fact, being
\begin{equation}\label{neq}
  n=\nonumber\frac{gT^3}{2\pi^2}\int_0^\infty \frac{x^2
  dx}{e^{\sqrt{x^2+(m/T)^2}}+1}\,,
\end{equation}
the function ${\cal F}$ becomes \cite{dolgov}
\begin{equation}\label{F}
  {\cal F}=2\left(\frac{m}{T}\right)^2HT^3\frac{d}{dy}\left[\frac{g}{2\pi^2}\int_0^\infty
\frac{x^2dx}{e^{\sqrt{x^2+y}}+1}\right]\,,
\end{equation}
where $y=(m/T)^2$ and it vanishes as $m=0$ \cite{dolgov,zee}.}
Similar results hold for anti-neutrinos:
\begin{equation}\label{n(T)-anti}
  n({\bar \nu})=\frac{gT^3}{2\pi^2}\int_{2 \sqrt{z}}^{\chi} dx
  \,\frac{x(\sqrt{x^2+z}+\sqrt{z})^2}{\sqrt{x^2+z}}\,\frac{1}{e^x+1}\,,
\end{equation}
and
\[
 {\cal F}({\bar \nu})=2\left(\frac{\lambda}{T}\right)^2HT^3
 \frac{d}{dz}\left[\frac{g}{2\pi^2}\int_{2\sqrt{z}}^{\chi} dx
  \,\frac{x(\sqrt{x^2+z}+\sqrt{z})^2}{\sqrt{x^2+z}}\,\frac{1}{e^x+1}\right]\,.
 \]
Thus the net neutrino asymmetry generated via Loop Quantum Gravity
effects would become
\begin{eqnarray}
 \Delta n &=& |n(\nu)-n({\bar \nu})|=\frac{2g\lambda T^2}{\pi^2}\int_0^{\chi}
  dx\, \frac{x}{e^x + 1}+\frac{gT^3}{2\pi^2}I(z) \nonumber \\
        &\approx& \frac{2g\lambda T^2}{\pi^2} \bigg[\frac{\pi^2}{12} +
        12 \sum_{n = 1}^\infty \, \frac{ (- \exp(\chi))^n}{n^2} +
        \frac{12}{{\cal L}  T} \, \ln(1 + e^{\chi}) - \frac{6}{({\cal L}  T)^2} \bigg]
        \,. \label{dnvalue}
\end{eqnarray}
where
\begin{equation}\label{I(z)}
  I(z)\equiv \int_0^{2\sqrt{z}}dx \,
  \frac{x(\sqrt{x^2+z}-\sqrt{z})^2}{\sqrt{x^2+z}(e^x+1)}\,.
\end{equation}
In evaluating (\ref{dnvalue}) we have neglected the contribution
coming from the $I(z)$-term since at low temperatures with respect
to Planck's one it is expected to be very small compared to other
terms. If we note that ${\cal L} \leq \lambda_D$ and use the upper
bound ${\cal L} \sim 1/{\bar p} \sim T$ where ${\bar p}$ is the de
Broglie momenta of neutrinos at a particular temperature, we can
estimate the neutrino asymmetry arising at that temperature.

Near GUT temperatures $T \sim 10^{16}$ GeV, the ratio of neutrino
asymmetry to entropy density, $\Delta n/s$ ($s\sim 0.44 g_* T^3$,
with $g_*\sim 10^2$ \cite{kolb}), turns out to be of the order of
$10^{-5}$ which would however be washed out by inflation.
Interesting temperatures would be near reheating temperatures
\footnote{In this case $\lambda/T \sim l_P/{\cal L}$ and the
corrections to the neutrino asymmetry due to $I(z)$ term would go
as $(l_P/{\cal L})^3$ which for the temperature range of $10^{11}$
GeV would be of the order of $10^{-24}$. Hence, our approximation
in Eq.(\ref{dnvalue}) is justified.} of the order of
$10^{10}-10^{11}$ GeV where this ratio would become of the order
of $10^{-10}$.  At lower temperatures the amount of asymmetry
generated would keep on decreasing till it becomes negligible,
though the asymmetry generated at reheating temperature would hold
till the neutrinos finally decouple. This lepton asymmetry would
lead to the baryon asymmetry through various GUT and electroweak
processes and thus contribute to the existing mechanisms to
produce matter-antimatter asymmetry in the Universe.

\section{Conclusion}

An intriguing prediction of modern approaches to quantum gravity
is a slight departure from Lorentz's invariance, which manifests
in a deformation of the dispersion relations of photons and
fermions. Such results have been indeed suggested in Loop Quantum
Gravity \cite{gambini,alfaroPRL,alfaroPRD}, String Theory
\cite{kostelecky,ellis} and Non-Commutative Geometry \cite{nonc}.
The former is endowed with a scale length characterizing the scale
on which new effects are non trivial, thus to wonder if there
exist different scenarios where these effects become testable (see
\cite{AlfaroPalma,neweffects}) is certainly of current interest.

In this letter we have shown that such modifications induced by
Loop Quantum Gravity might help to put some light on  unsolved
problems like matter-antimatter asymmetry in Standard Model.
Application of weave states for Majorana fermions naturally leads
to difference in energies for different chiralities which may be
interpreted as difference in particle and antiparticle energies
for the case of massless neutrinos. This leads to asymmetry
between matter and antimatter species and yields the observed
value at around reheating temperatures.Our proposal introduces a
way for generation of matter-antimatter asymmetry via Loop Quantum Gravity,
whose complete analysis would require
relaxing the massless limit and secondly taking into account
various standard model interactions in unison with Loop Quantum Gravity.
Then we shall be able to know how the above mechanism to generate
matter-antimatter asymmetry contributes relative to other processes.
This opens up a new arena to make phenomenological studies in Loop
Quantum Gravity in future.

It is a remarkable phenomena that quantum structure of spacetime
itself may generate matter-antimatter asymmetry in the Universe.
In fact, this might be a generic feature of theories of Quantum
Gravity. It reflects that Quantum Gravity may lead to effects
occurring at lower energy scales, specially in the desert between
electroweak and Planck scale, which may provide natural answers to
some unsolved problems.


\acknowledgments P.S. thanks Council for Scientific and Industrial
Research for a research grant. The authors thank the referee for
constructive comments.

\end{document}